\documentclass[aps,epsfig,prb,unsortedaddress,preprint,longbibliography]{revtex4-1}

\usepackage{amsmath}
\usepackage{amsfonts}
\usepackage{graphicx}

\def\nn{\nonumber}

\begin{document}

\title{Band Structures of Edge Magnetoplasmon Crystals}

\author{Ken-ichi Sasaki}
\email{kenichi.sasaki.af@hco.ntt.co.jp}
\affiliation{NTT Research Center for Theoretical Quantum Physics and NTT Basic Research Laboratories, NTT Corporation,
3-1 Morinosato Wakamiya, Atsugi, Kanagawa 243-0198, Japan}


\date{\today}
 
\begin{abstract}
 A two-dimensional electron gas in a static external magnetic field 
 exhibits two distinct collective excitation modes.
 The lower frequency mode propagates along the periphery of the domain 
 almost freely with an extended lifetime, which is referred to as edge magnetoplasmons.
 Peculiar phenomena caused by a capacitive interaction between nearest neighbor domains are known, 
 such as the emergence of Tomonaga-Luttinger liquid and charge density fractionalization.
 Meanwhile, the number of coupled domains investigated in the past has been limited to a small number.
 Here, we performed calculations using a continuum model of edge magnetoplasmons,
 the band structures of planar crystals composed of an arbitrary number of domains, 
 including a chain, ladder, and honeycomb network, with the general interaction strength.
 We explain the band structures in terms of the fundamental collective modes of a molecule 
 composed of two equivalent domains.
 These are the extended chiral propagation modes that yield a linear dispersion band
 and the standing wave modes localized in the coupled regions that cause a flat band.
 The chain's band structures resemble the miniband structures
 calculated from the Kronig-Penny model for the electron in a semiconductor superlattice.
 We point out that a geometrical deformation of the chain does not change the band structures as 
 it can be expressed as a gauge degree of freedom that only causes a shift in the wavenumber.
\end{abstract}

\pacs{}
\maketitle

\section{Introduction}

Many physical systems have the collective excited states, known as plasmons,
in which the electrons and electromagnetic field are dynamically coupled to 
form a self-sustainable motion of the composites.
When plasmons exist in each component (or domain),
they can interact with each other through electromagnetic fields
if the domains are close enough to each other.
In this manuscript, we examine the characteristics of the plasmons of the entire system,
as regards the basic plasmons localizing and propagating along the edge of each planar domain,
which are referred to as the edge magnetoplasmons (EMPs).~\cite{Allen1983,Mast1985,Glattli1985}
EMPs are the low-energy excited states of a planar system of a two-dimensional electron gas
in a stationary external magnetic field applied perpendicular to the plane,
and they exhibit a chiral propagation that moves in a direction determined by 
the orientation of the magnetic field.
It is known that EMPs exhibit peculiar phenomena caused by a capacitive interaction between nearest neighbor 
domains, such as the emergence of Tomonaga-Luttinger liquid and 
charge density fractionalization.~\cite{Hashisaka2018a}
There is also a theoretical proposal that 
EMPs are potential candidates for quantum energy teleportation, in which 
energy transportation can be realized by classical information without energy carriers, 
and the interaction between EMPs plays an essential role in it.~\cite{Yusa2011}

Besides the fact that experiments are scheduled and theoretical consideration is called for,
we have other motivation for getting onto the subject of the EMPs in a domain network.
First, if we regard a single planar domain, at the edge of which EMPs exist, as a fictitious atom, 
our objective is to find the energy spectrum of a plasmonic crystal, or more specifically,
an EMP crystal or EMP molecule.
Naturally, since a plasmon is a hybrid of electrons and photons (electromagnetic fields),
such a plasmonic crystal must have an essential relationship to a photonic crystal.
Indeed, we will show that there is a close similarity between an EMP crystal and 
semiconductor superlattice, which is a periodic structure of layers of two (or more) materials
and the simplest example of a photonic crystal.
When discussing the interaction between adjacent domains, 
the idea of static atomic orbitals, such as the bonding and anti-bonding orbitals
which are useful in discussing the formation or stability of a lattice, 
may be extended to the chiral and dynamical counterparts.
Second, we seek to gain a better understanding of the nature of EMPs in a network of domains.
For example, we would like to know 
the lowest energy excited state of the whole system.
Is it still an EMP that propagates along the outer edge of the whole system?

We propose in this manuscript a general method to calculate the energy spectrum of a planar EMP crystal.
We first show the results for the simplest EMP molecule consisting of the two domains (Sec.~\ref{sec:twodomains}), 
which are applicable for any finite number of domains and essential to understand the physics.
Next, the dispersion relation of the plasmons in the periodic system consisting of $N$ domains, 
namely the energy band structures of the periodic EMP crystal, is analytically constructed 
for a chain, ladder, and honeycomb network (Sec.~\ref{sec:periodicdomains}).
We will show that the naive ``EMP'' of the EMP crystal, which has the same chirality as the atomic EMP, 
is not the lowest energy state of the entire system for the general interaction strength.
We also discuss an extension of the planar EMP crystals toward three-dimensional counterparts,
which is useful in identifying the topological aspect of a system.

\section{Basic Knowledge about EMP in a Single Domain}

In this section, 
we review the two main properties of an EMP in a single domain 
and introduce an effective model used for our analysis.

\subsection{Two Main Properties of EMP}

First, an EMP pulse propagates almost freely
along the edge of a two-dimensional electron gas 
in the direction determined by the magnetic-field orientation.~\cite{Ashoori1992}
This suggests that the dispersion relation of EMPs is approximately linear.
Indeed, when the boundary potential is sufficiently sharp 
so that the electron density changes abruptly,
Volkov and Mikhailov 
solved an integral equation of the electric potential with the Wiener-Hopf method under reasonable assumptions
and succeeded in getting the dispersion relation as~\cite{volkov88,SergeiA.Mikhailov2001}
\begin{align}
 \omega(q_y) = \frac{2q_y \sigma_{xy}}{\kappa}
 \left( \ln\frac{2}{|q_y|\ell_x} + 1 \right),
\label{eq:VM}
\end{align}
where $q_y$ is the wavevector along the edge, 
$\sigma_{xy}$ is the static Hall conductivity, $\kappa$ is the relative dielectric constant,
and $\ell_x$ is the localization length of the charge density (in the direction perpendicular to the edge),
which is proportional to the dynamical conductivity $\sigma_{xx}(\omega)$:
\begin{align}
 \ell_x = \frac{2\pi i \sigma_{xx}(\omega)}{\omega \kappa}.
\label{eq:ell}
\end{align}
Since $\ell_x$ may depend on $\omega$, 
the EMP frequency is determined by solving Eqs.~(\ref{eq:VM}) and (\ref{eq:ell}) self-consistently.
Practically, $\ell_x$ is independent of $\omega$ as $\ell_x \simeq \frac{e^2\nu}{\kappa \hbar \omega_c}$
(where $\nu$ is the filling factor, and $\omega_c$ is the cyclotron frequency), 
because $\sigma_{xx}(\omega) \simeq -i \frac{\omega}{2\pi} \frac{e^2 \nu}{\hbar \omega_c}$ holds
and $\omega$ in the numerator of Eq.~(\ref{eq:ell}) is canceled out by that of the denominator.
The wavelength of interest is usually much larger than $\ell_x$,
which makes the dispersion relation of EMPs approximately linear.
The linear chiral dispersion is in sharp contrast to the gapped spectrum of bulk magnetoplasmons (MPs), 
which is written in terms of $\omega_c$ and two-dimensional plasmon frequency $\omega_p$
as $\sqrt{\omega_p^2 + \omega_c^2}$.

Second, the EMP damping is suppressed by the applied magnetic field.
If damping is significant, EMPs would not be observed in a strong magnetic field,
because $\sigma_{xy}$ makes the frequency lower and lower by increasing the magnetic field
and the Drude peak may obscure the EMP signal.
The origin of the long EMP lifetime is a subtle problem.
In a previous paper,~\cite{Sasaki2016}
we argued that
an internal magnetic field was neglected in theoretical approaches~\cite{volkov88} and 
that this simplification prevented the EMP lifetime from being determined.
On the other hand, we found that 
the following approximate relationship between the EMP lifetime and MPs exists:
\begin{align}
 \tau_{\rm emp} = \frac{\omega_c^2+\omega_p^2}{\omega_p^2}\tau_{\rm mp}.
 \label{eq:tauemp}
\end{align}
This result is obtained by noticing that peculiar plasmons
whose frequencies are purely imaginary exist
in the interior of a two-dimensional electron gas described by the Drude model.~\cite{falko1989} 
When an external magnetic field is applied to the system,
these bulk plasmons are still non-oscillating and are isolated from the MP. 
They are mainly in a transverse magnetic mode and can
combine with a transverse electronic mode locally at an edge of the
system to form EMPs.
We note that though Eq.~(\ref{eq:tauemp}) reasonably explains experimental results,~\cite{Yan2012}
the derivation is classical, and that whether it can be extended to the quantum Hall regime is unknown.

The quantum Hall effect (QHE) is not the necessary condition for the existence of EMPs, 
but EMP lifetime is elongated by the QHE.
The QH state is characterized by an electronic ground state 
whose excitation spectrum is gapful, an incompressible liquid state,
in the bulk but is gapless at the edge.
The energy spectrum of the QHE is similar to that of MPs, 
namely an MP is gapful, but an EMP is gapless.
The dynamical aspect of the edge states in fractional QHEs,
where interactions between electrons contribute to an incompressible state,
has been explored by many authors.~\cite{MacDonald1990a,Wassermeier1990,Wen1990,Wen1990a,WEN1992,Kane1994a,Ezawa2013}

\subsection{One-dimensional effective model}

The results for a single domain presented above, 
which are based on the classical field theory of electrodynamics, 
are essential and very useful in understanding experimental results.
However, they are difficult to extend to more complicated physical circumstances
in which EMPs interact with each other.
The presence of $\kappa$ in Eq.~(\ref{eq:VM}) already suggests that 
the propagation velocity of EMPs depends on the system parameters,
including its environments.~\cite{Kumada2020}
Hashisaka {\it et al}.~\cite{Hashisaka2013} 
proposed a distributed-element circuit model of interacting EMPs, 
which introduces a geometrical capacitance $c_x$ 
for mutual interactions in addition to a channel capacitance that simulates
the propagation velocity of an isolated EMP as
\begin{align}
 v = \frac{\sigma_{xy}}{c_{ch}}.
 \label{eq:Vch}
\end{align}
This model is plausible and capable of describing the capacitive interactions 
between counter-propagating EMPs,~\cite{Hashisaka2013,Kamata2014}
as well as those between co-propagating EMPs.~\cite{Hashisaka2017,Hashisaka2018a}
In the coupled region of the two domains, the chiral nature of the EMPs in each domain 
is disturbed by $c_x$, and the mixed mode is formed as a non-chiral standing wave.
The model can be extended to include the effect of a general type of gate needed 
to control the carrier density or the velocity.~\cite{Kumada2020}
We therefore adopt the model in calculating the energy spectrum of coupled domains.

\section{Two Domains}\label{sec:twodomains}

\begin{figure}[htbp]
 \begin{center}
  \includegraphics[scale=0.8]{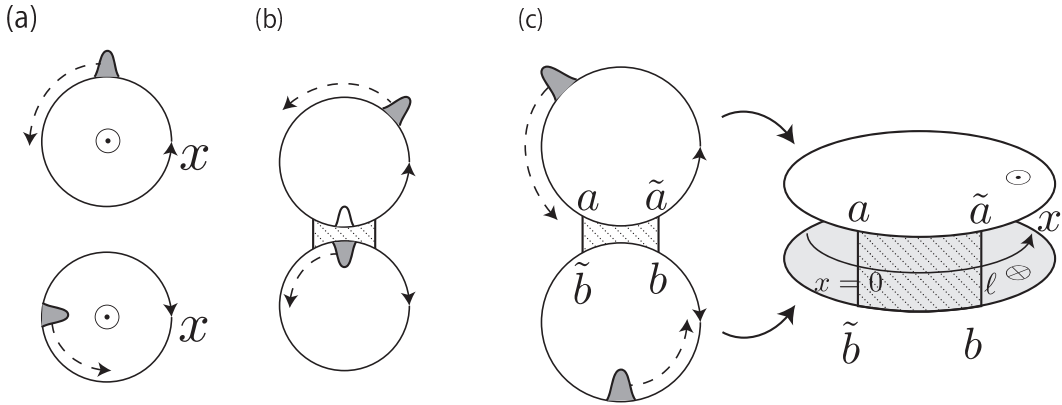}
 \end{center}
 \caption{{\bf Geometries of the two domains.}
 (a) When the two domains are independent, EMPs are freely propagating along each edge.
 (b) When the two domains are close enough to each other, 
 they interact with each other in the shaded part through inter-edge capacitive coupling.
 (c) The spatial coordinate $x$ may be shared by the two domains 
 through the procedure discussed in the main text.
 }
 \label{fig:emp_coupled}
\end{figure}

We assume that when the interaction between the two domains is negligibly small,
the EMP pulses [expressed by humps in Fig.~\ref{fig:emp_coupled}(a)]
can propagates independently along the edge of each domain with velocity $v$
and without any dissipation.
We will neglect the complicated EMP profile in the direction perpendicular to the edge
and focus on the dynamics along the edge.
The EMP dynamics of the first domain is expressed by the normal modes of the current and charge densities, 
$j_1(x,t) = a v e^{i\frac{\omega}{v}(x-vt)}$ and $\rho_1(x,t) = a e^{i\frac{\omega}{v} (x-vt)}$,
which are a function of $x-vt$ that represents the chiral character of EMPs.
The charge density (at $x=vt$) can be positive or negative depending on the sign of $a$.
A positive and negative current density means a positive and negative charge density, respectively, 
that propagates in the same direction determined by the chirality.
The continuity equation expressing charge conservation in the first domain is given by
$\partial_t \rho_1(x,t) +\partial_x j_1(x,t) =0$.
Likewise, we define the normal modes of 
current and charge densities for the second domain as
$j_2(x,t) = b v e^{i\frac{\omega}{v} (x-vt)}$ and 
$\rho_2(x,t) = b e^{i\frac{\omega}{v}(x-vt)}$.
The eigenfrequencies are quantized by the periodic boundary condition as
$\omega_1 = (2\pi v /L_1)n$ and $\omega_2 = (2\pi v/L_2)n$ with integer $n$,
where $L_1$ and $L_2$ are circumference of the first and second domains, respectively.

The direction of spatial coordinate $x$ is not necessarily the same 
(for example, anticlockwise) for the two domains.
Rather, when we consider the effects of coupling between the two domains,
it turns out to be convenient to define the coordinate for the second domain 
in the direction opposite to that for the first domain.
In this new coordinate system, we have 
$j_2(x,t) = -b v e^{-i\frac{\omega}{v} (x+vt)}$ and 
$\rho_2(x,t) = b e^{-i\frac{\omega}{v} (x+vt)}$, by the replacement $v\to -v$.
The minus sign is added to $j_2$ only (not in front of $b$ of $\rho_2$), 
which is necessary for them to satisfy the continuity equation 
$\partial_t \rho_2(x,t) +\partial_x j_2(x,t) =0$.
Note that 
$j_2(x,t)$ and $\rho_2(x,t)$ become a function of $x+vt$, 
showing the same chirality as the EMPs in the first domain.

The sign difference between $j_1$ and $j_2$ (in front of $b$) may be explained
by a fictitious procedure in three dimensions, in which 
the second domain is turned inside out and placed below (or above) the first domain,
as shown in Fig.~\ref{fig:emp_coupled}(c).
Note that the orientation of the second domain is reversed and that 
the direction of a magnetic field normal to the second domain plane is reversed too.
As a result, we can regard the total system 
as if the Hall conductivities for the two domains have different signs;
\begin{align}
 \begin{pmatrix}
  j_1(x) \cr
  j_2(x)
 \end{pmatrix}
 = \sigma_{xy}
 \begin{pmatrix}
  1 & 0 \cr 
  0 & -1
 \end{pmatrix}
 \begin{pmatrix}
  V_1(x) \cr V_2(x)
 \end{pmatrix},
 \label{eq:jh}
\end{align}
where $V_i(x)$ is the EMP potential given by $V_i(x) = \frac{\rho_i(x)}{c_{ch}}$.
Indeed, using Eq.~(\ref{eq:Vch}), we show that $j_1(x,t)=v \rho_1(x,t)$ and $j_2(x,t)=-v \rho_2(x,t)$,
which are consistent with the normal modes.
The fictitious procedure in three-dimensions makes us to notice that 
this system is topologically not equivalent to a capacitor in an external magnetic field
(rather it is equivalent to a capacitor containing a magnetic monopole).
Meanwhile, there is an EMP molecule with a staggered magnetic field 
that corresponds to a capacitor in a magnetic field, which is discussed in Appendix~\ref{app:sd}.

When the two domains are sufficiently close, 
they couple with each other through a capacitive coupling $c_x$
in the region $x\in [0,\ell]$ represented by shaded part between the two domains in Fig.~\ref{fig:emp_coupled}(b).
We assume that $c_x$ is a constant in the coupled region and vanishes outside.
The capacitive coupling modifies charge densities through 
a difference between the EMP potentials of the two domains as
$\rho_1(x) = c_{ch} V_1(x) + c_{x}(V_1(x)-V_2(x))$
and $\rho_2(x) = c_{ch} V_2(x) + c_x(V_2(x)-V_1(x))$.
These are expressed with a $2\times 2$ matrix as
\begin{align}
 \begin{pmatrix}
  V_1(x) \cr V_2(x)
 \end{pmatrix}
 = \frac{1}{c_{ch}}
 \begin{pmatrix}
  1-\delta & \delta \cr
  \delta & 1-\delta
 \end{pmatrix}
 \begin{pmatrix}
  \rho_1(x) \cr
  \rho_2(x)
 \end{pmatrix},
 \label{eq:vrho}
\end{align}
by defining a coupling constant 
\begin{align}
 \delta \equiv \frac{c_x}{c_{ch}+2c_x}.
\end{align}
By combing Eqs.~(\ref{eq:vrho}) and (\ref{eq:jh}), we have
\begin{align}
 \begin{pmatrix}
  j_1(x) \cr
  j_2(x)
 \end{pmatrix}
 =v
 \begin{pmatrix}
  1-\delta & \delta \cr
  -\delta & -(1-\delta)
 \end{pmatrix}
 \begin{pmatrix}
  \rho_1(x) \cr
  \rho_2(x)
 \end{pmatrix}.
 \label{eq:jrho}
\end{align}
Because of the continuity equation expressing independent charge conservation in each domain 
$\partial_t \rho_i(x,t) +\partial_x j_i(x,t) =0$,
Eq.~(\ref{eq:jrho}) becomes the following dynamical equation of the current density
\begin{align}
 \partial_t
 \begin{pmatrix}
  j_1(x,t) \cr
  j_2(x,t)
 \end{pmatrix}
 =-v
 \begin{pmatrix}
  1-\delta & \delta \cr
  -\delta & -(1-\delta)
 \end{pmatrix}
\partial_x
 \begin{pmatrix}
  j_1(x,t) \cr
  j_2(x,t)
 \end{pmatrix}.
 \label{eq:jc}
\end{align}
The eigenvalues of the $2\times 2$ matrix are $\pm v_c$, 
where $v_c \equiv v\sqrt{1-2\delta}$ corresponds to the propagation velocity in the coupled region,
which is slower than that in the uncoupled region ($v=\sigma_{xy}/c_{ch}$) since $\delta \ge 0$.
The EMP in the coupled region is not chiral;
there are modes propagating in the forward (or right) and backward (or left) directions along the $x$ axis.
By expanding current density using eigenspinors of the $2\times 2$ matrix,
we have for $x \in [0,\ell]$
\begin{align}
 \begin{pmatrix}
  j_1(x) \cr
  j_2(x)
 \end{pmatrix}
 = 
 \alpha_R 
 \begin{pmatrix}
  1 \cr -r 
 \end{pmatrix}
 e^{+i\frac{\omega}{v_c} x}
 -
 \alpha_L
 \begin{pmatrix}
  -r \cr 1
 \end{pmatrix}
 e^{-i\frac{\omega}{v_c} x}, 
\label{eq:jc}
\end{align}
where 
\begin{align}
 r \equiv \frac{1-\delta-\sqrt{1-2\delta}}{\delta}, \ \ 
 \left( v_c = \frac{1-r}{1+r} v \right).
\end{align}
Because $\delta$ is an increasing function of $c_x$ with upper bound $1/2$, 
we define the weak and strong coupling limit as 
$\delta \to 0$ and $1/2$ (or $r\to 0$ and $1$), respectively. 
The first term on the right-hand side of Eq.~(\ref{eq:jc})
represents the mode propagating with the positive velocity in the coordinate $x$ with amplitude $\alpha_R$.
Using the continuity equation, or by substituting Eq.~(\ref{eq:jc}) into Eq.~(\ref{eq:jrho}), 
we obtain the charge density
\begin{align}
 \begin{pmatrix}
  \rho_1(x) \cr
  \rho_2(x)
 \end{pmatrix}
 = \frac{1}{v_c} \left\{
 \alpha_R 
 \begin{pmatrix}
  1 \cr -r 
 \end{pmatrix}
 e^{+i\frac{\omega}{v_c} x}
 +
 \alpha_L
 \begin{pmatrix}
  -r \cr 1
 \end{pmatrix}
 e^{-i\frac{\omega}{v_c} x} \right\}.
 \label{eq:rho}
\end{align}

Next, we examine the boundary conditions to be satisfied for
the boundaries of the coupled region at $x=0$ and $\ell$.
By the spatial integration of the continuity equation 
over an infinitesimal region including the boundary, 
it is shown that the current must be continuous there;
\begin{align}
 \lim_{\epsilon\to 0}
 \int_{x'-\epsilon}^{x'+\epsilon} dx \partial_x j_1(x,t) =
 - \partial_t \int_{x'-\epsilon}^{x'+\epsilon} dx \rho_1(x,t) 
 \to
 j_i(x'+0) = j_i(x'-0).
\end{align}
Therefore, by setting $j_1(0)=a$, $j_1(\ell)=\tilde{a}$, $j_2(0)=-\tilde{b}$, 
and $j_2(\ell)=-b$, we obtain from Eq.~(\ref{eq:jc})
\begin{align}
 & a = \alpha_R + r \alpha_L, \ \ -\tilde{b} = -r \alpha_R - \alpha_L, \nn
 \\
 & \tilde{a} = \alpha_R e^{i\frac{\omega}{v_c}\ell} + r \alpha_L e^{-i\frac{\omega}{v_c}\ell}, \ \ 
 -b = -r \alpha_R e^{i\frac{\omega}{v_c}\ell} - \alpha_L e^{-i\frac{\omega}{v_c}\ell}.
 \label{eq:bc}
\end{align}
We note that the charge density is not continuous at the boundaries.
Such a discontinuity is easy to recognize by considering a square wave of width $\Delta x$ 
as an incident wave prepared in the uncoupled region.
When it enters the coupled region, the width must decrease to $\frac{v_c}{v}\Delta x$
and the charge density must increase because of the charge conservation.
Even though the discontinuity of the charge density, by itself, does not result in any serious error,
it might represent poor modeling on the boundary.
Indeed, according to Volkov's theory, $\ell_x$ actually depends on $\kappa$, 
so $\ell_x$ may be changed at the boundary.
There is a possibility that a charge flow in the direction perpendicular to the edge 
may exist at the boundary.
In this manuscript,
we ignored the possible effect due to the discontinuous change in the charge density.

By eliminating $\alpha_R$ and $\alpha_L$ from the above equations,
we get a $2\times 2$ symplectic (transfer) matrix with a unit determinant
that relates the current density of one domain to that of the other domain as
\begin{align}
 \begin{pmatrix}
  a \cr \tilde{a}
 \end{pmatrix}
 =
 T(\omega)
 \begin{pmatrix}
  \tilde{b} \cr b
 \end{pmatrix},
 \label{eq:mat}
\end{align}
where
\begin{align}
 T(\omega) 
 \equiv
 \begin{pmatrix}
  1 & 1 \cr
  e^{i\frac{\omega}{v_c}\ell} & e^{-i\frac{\omega}{v_c}\ell}
 \end{pmatrix}
 \begin{pmatrix}
  \frac{1}{r} & 0 \cr
  0 & r
 \end{pmatrix}
 \begin{pmatrix}
  1 & 1 \cr
  e^{i\frac{\omega}{v_c}\ell} & e^{-i\frac{\omega}{v_c}\ell}
 \end{pmatrix}^{-1}
 = \frac{1}{-2i r \sin\left(\frac{\omega \ell}{v_c} \right)}
 \begin{pmatrix}
  t_\omega & -t_0 \cr
  t_0 & -t_\omega^*
 \end{pmatrix},
 \label{eq:Tmat}
\end{align}
and $t_\omega \equiv e^{-i\frac{\omega}{v_c}\ell} - r^2 e^{+i\frac{\omega}{v_c}\ell}$
(and therefore $t_0 = 1 - r^2$).
Because an EMP propagates freely in the uncoupled region of each domain, 
we have a phase relationship between $a$ ($b$) and $\tilde{a}$ ($\tilde{b}$) as follows:
\begin{align}
 \begin{cases}
  & e^{+i\frac{\omega (L_1-\ell)}{v}} \tilde{a} = a, \\
  & e^{+i\frac{\omega (L_2-\ell)}{v}} \tilde{b} = b.  
 \end{cases}
 \label{eq:tb0}
\end{align}
Substituting Eq.~(\ref{eq:tb0}) into Eq.~(\ref{eq:mat}),
we obtain
\begin{align}
 \begin{pmatrix}
  1 \cr e^{-i\frac{\omega (L_1-\ell)}{v}} 
 \end{pmatrix}
 a
 = T(\omega)
 \begin{pmatrix}
  e^{-i\frac{\omega (L_2-\ell)}{v}} \cr 1
 \end{pmatrix}
 b.
 \label{eq:ab}
\end{align}
By multiplying $(1,-e^{i\frac{\omega (L_1-\ell)}{v}})$ with the both sides of Eq.~(\ref{eq:ab}),
we obtain the equation written as
\begin{align}
 {\rm Re} \left[ e^{-i\frac{\omega}{v}\left( \frac{L_1+L_2}{2}-\ell \right)} t_\omega \right]
 = t_0 \cos \left( \frac{\omega}{2v}\left( L_1-L_2 \right) \right),
\end{align}
which determines the possible eigenfrequencies.
This is simplified when the two domains are geometrically equivalent, i.e., $L_1=L_2\equiv L$, as
\begin{align}
 r^2 \sin^2\left( \frac{\omega (L-\ell)}{2v}- \frac{\omega \ell}{2v_c}\right) 
 = \sin^2 \left( \frac{\omega (L-\ell)}{2v}+ \frac{\omega \ell}{2v_c} \right).
 \label{eq:link2}
\end{align}
This equation can be solved numerically in general and analytically in a certain limit.

\begin{figure}[htbp]
 \begin{center}
  \includegraphics[scale=0.8]{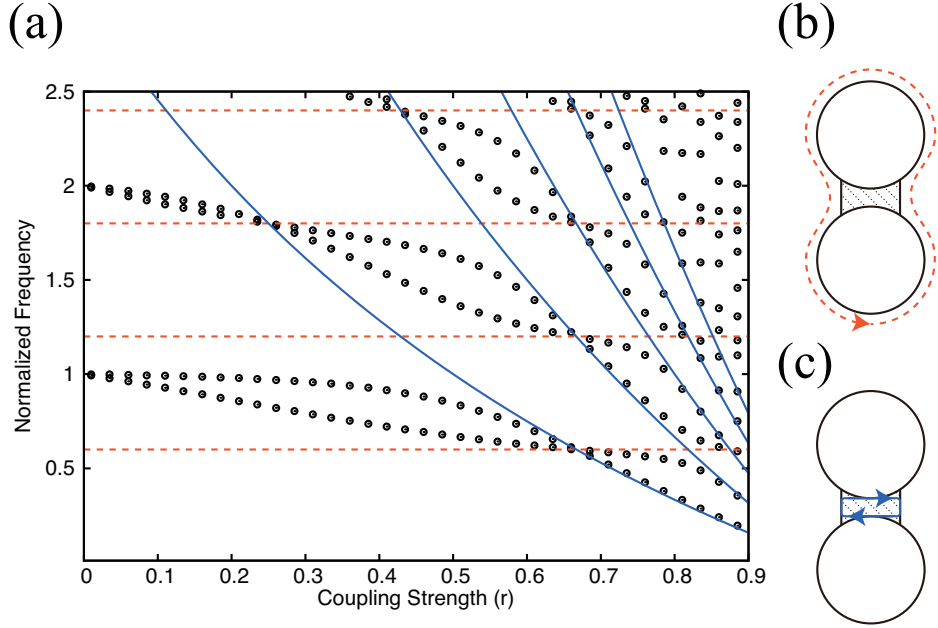}
 \end{center}
 \caption{{\bf Low-energy spectrum of a simplest EMP molecule.}
 (a) The calculated frequency is plotted as dots in units of the fundamental frequency of an isolated domain 
 without the interaction ($\frac{2\pi v}{L}$).
 This result is for $L=6\ell$.
 The dashed lines represent new EMP modes ($\omega_e$, $2\omega_e$, $3\omega_e$ and $4\omega_e$) 
 that propagate along the outer edge (b),
 and the solid curves represent the localized standing wave modes 
 ($\omega_s$, $2\omega_s$, $3\omega_s$, $4\omega_s$, and $5\omega_s$) in the coupled region (c).
 }
 \label{fig:2link}
\end{figure}

Figure~\ref{fig:2link}(a) shows the numerical solution of Eq.~(\ref{eq:link2}) for $L=6\ell$
as a function of the coupling strength ($r$).
The interaction always decreases the frequency.
In the weak coupling limit,
there are two fundamental modes with equal angular frequency $\omega=2\pi v/L$.
The energies of the originally degenerate states are split and cross again (at $r\simeq 0.66$)
by increasing capacitive coupling.
The possible crossing points and behavior in a strong coupling regime
can be understood on physical grounds, by introducing the following two modes. 
One physically expected mode has the fundamental frequency 
\begin{align}
 \omega_e = \frac{\pi v}{L-\ell},
\end{align}
which corresponds to a new EMP mode moving around the periphery of the two coupled domains
[see Fig.~\ref{fig:2link}(b)].
The eigenfrequency of the other mode is a multiple of
\begin{align}
 \omega_s = \frac{\pi v_c}{\ell},
\end{align}
which represents a standing wave localized in the coupled region
[see Fig.~\ref{fig:2link}(c)].
This becomes the lowest frequency mode in the strong coupling limit, 
while it is a high-frequency mode in weak coupling.
Please note that the $T(\omega)$ matrix is ill-defined exactly when $\omega=n\omega_s$ and that
Eq.~(\ref{eq:bc}) gives $\tilde{a}=a(-1)^n$ and $\tilde{b}=b(-1)^n$, 
which are inconsistent with the phase condition of Eq.~(\ref{eq:tb0}).
Thus, even for a strong coupling case, 
the calculated frequencies in Fig.~\ref{fig:2link}(a)
are very slightly displaced from $n \omega_s$.

The level crossing between $\omega_s$ and $\omega_e$ occurs when $r = 1- \frac{2\ell}{L} (\equiv r_c)$.
The critical coupling strength is determined by the geometrical parameters $\ell$ and $L$ only.
The spectrum at the critical point exhibits a special feature that the possible frequencies 
are exact multiples of the fundamental frequency.

By multiplying $(1,e^{i\frac{\omega (L-\ell)}{v}})$ with the both sides of Eq.~(\ref{eq:ab}), 
we know that 
\begin{align}
 \frac{a}{b} 
 = \frac{1+r^2}{2r}\cos\left(\frac{\omega(L-\ell)}{v}\right)
 + \frac{1-r^2}{2r}
 \sin\left(\frac{\omega(L-\ell)}{v}\right)\cot \left(\frac{\omega \ell}{v_c}\right)
 \label{eq:ab_2rings}
\end{align}
holds for the general value of $\omega$.
Meanwhile, solutions of Eq.~(\ref{eq:link2}) satisfy either 
\begin{align}
 -r \sin\left( \frac{\omega (L-\ell)}{2v}- \frac{\omega \ell}{2v_c}\right) 
 = \sin \left( \frac{\omega (L-\ell)}{2v}+ \frac{\omega \ell}{2v_c} \right)
 \label{eq:sym}
\end{align}
or 
\begin{align}
 r \sin\left( \frac{\omega (L-\ell)}{2v}- \frac{\omega \ell}{2v_c}\right) 
 = \sin \left( \frac{\omega (L-\ell)}{2v}+ \frac{\omega \ell}{2v_c} \right).
 \label{eq:asym}
\end{align}
It is shown by combining Eqs.~(\ref{eq:ab_2rings}) and (\ref{eq:sym}) or (\ref{eq:asym}) 
that $a/b$ must be $+1$ or $-1$ in order that the solutions exist for the general coupling strength.
In the weak coupling limit, the higher (lower) frequency state has $a/b=+1$ ($-1$).
The higher or lower frequency characteristics change when the two modes cross each other 
with increasing $r$.

Putting $a/b=+1$ ($-1$) into Eq.~(\ref{eq:bc}),
we obtain $\alpha_R=e^{-i\frac{\omega}{v_c}\ell} \alpha_L$ ($\alpha_R=-e^{-i\frac{\omega}{v_c}\ell} \alpha_L$), 
by which Eqs.~(\ref{eq:jc}) and (\ref{eq:rho}) are determined with the exception of the normalization factor.
The current and charge densities for $a/b=+1$ are
\begin{align}
 &
 \begin{pmatrix}
  j_1(x) \cr
  j_2(x)
 \end{pmatrix}_{+1}
 = 
 \alpha_L e^{-i\frac{\omega \ell}{2v_c}}
 \begin{pmatrix}
  (1+r)\cos \left(\frac{\omega}{v_c}(x-\frac{\ell}{2}) \right) 
  + i(1-r) \sin \left(\frac{\omega}{v_c}(x-\frac{\ell}{2}) \right) 
  \cr
  -(1+r)\cos \left(\frac{\omega}{v_c}(x-\frac{\ell}{2}) \right) 
  + i(1-r) \sin \left(\frac{\omega}{v_c}(x-\frac{\ell}{2}) \right) 
 \end{pmatrix},
 \\
 &
 \begin{pmatrix}
  \rho_1(x) \cr
  \rho_2(x)
 \end{pmatrix}_{+1}
 = 
 \frac{\alpha_L e^{-i\frac{\omega \ell}{2v_c}}}{v_c}
 \begin{pmatrix}
  (1-r)\cos \left(\frac{\omega}{v_c}(x-\frac{\ell}{2}) \right) 
  + i(1+r) \sin \left(\frac{\omega}{v_c}(x-\frac{\ell}{2}) \right) 
  \cr
  (1-r)\cos \left(\frac{\omega}{v_c}(x-\frac{\ell}{2}) \right) 
  - i(1+r) \sin \left(\frac{\omega}{v_c}(x-\frac{\ell}{2}) \right) 
 \end{pmatrix}.
\end{align}
The current and charge densities for $a/b=-1$ are given 
by exchanging current with charge for $a/b=+1$ as 
\begin{align}
 &
 \begin{pmatrix}
  j_1(x) \cr
  j_2(x)
 \end{pmatrix}_{-1}
 = -v_c
 \begin{pmatrix}
  \rho_1(x) \cr
  \rho_2(x)
 \end{pmatrix}_{+1},
 \ \
 \begin{pmatrix}
  \rho_1(x) \cr
  \rho_2(x)
 \end{pmatrix}_{-1}
 = -\frac{1}{v_c}
 \begin{pmatrix}
  j_1(x) \cr
  j_2(x)
 \end{pmatrix}_{+1}.
\end{align}
Since sine terms vanish at the center of the coupled region ($x=\ell/2$) for any $\omega$,
we first assume the convention that the normalization factor of $\alpha_L e^{-i\frac{\omega \ell}{2v_c}}$
is a real number.
Then the signs of $a/b$ represent different configurations of the dipole moments.
For $a/b=+1$,
the charge densities at the two domains in the coupled region have the same sign 
(like ``anti-bonding orbital''). 
The direction of the dipole moment in each domain points in the opposite direction, 
and a net dipole moment of the two domains vanishes in total.
Meanwhile, for $a/b=-1$, 
the charge densities at the two domains in the coupled region have different signs 
(like ``bonding orbital'').
The direction of the dipole moment of each domain points in the same direction,
and the two domains constructively make a large dipole moment as a whole.
The above convention is invalid unless $\alpha_L e^{-i\frac{\omega \ell}{2v_c}}$ is a real number,
because the normalization factor is a complex number in general.
For example, in a strong coupling region, $\omega = n \omega_s$ and 
the normalization factor of $\alpha_L (-i)^n$, where $\alpha_L$ is a real number.
The dipole moment characteristics change as $r$ increases.
Generally, we can specify the phase of $\alpha_L$ for a given $\omega$,
because $[j_1(0)]_{\pm 1}=(\pm r + e^{-i\frac{\omega \ell}{v_c}})\alpha_L$ is a real number.

We consider a geometrical case of $L \gg \ell$ in which 
an incident steady current flows in the first domain towards the coupled region.
This situation is expressed by setting $a=1$ in Eq.~(\ref{eq:mat}).
Because the EMP of the second domain propagates in the counterclockwise direction
as shown in Fig.~\ref{fig:emp_coupled}(b),
it takes a very long time to arrive at $x=\ell$ from $x=0$.
We therefore may assume that $b=0$ in Eq.~(\ref{eq:mat}).
From these conditions, we obtain the reflectance and transmittance as
\begin{align}
 R \equiv |\tilde{a}|^2 = \left|
 \frac{(1-r^2) e^{i\frac{\omega \ell}{v_c}}}{1-r^2 e^{2i\frac{\omega \ell}{v_c}}} \right|^2,
 \ \ 
 T \equiv |\tilde{b}|^2 = \left|
 \frac{r (1-e^{2i\frac{\omega \ell}{v_c}})}{1-r^2 e^{2i\frac{\omega \ell}{v_c}}} \right|^2.
\end{align}
This result coincides with the result
known for the reflection and transmission of light by thin films.~\cite{Palik1970,Heavens1960}
The coupled region can be expressed as a non-absorbing medium 
with the refractive index of $n\equiv v/v_c$ or $n = (1+r)/(1-r)$.
When the frequency of an incident wave matches 
the frequency of a standing wave (i.e. when $\omega$ is a multiple of $\omega_s$),
perfect reflection with $R=1$ and $T=0$ is realized.
In the strong coupling limit,
nearly perfect transmission is expected when $\omega = \left(n+\frac{1}{2} \right)\omega_s$,
where $n=0,1,\cdots$.

\section{Periodic domains}\label{sec:periodicdomains}

In this section,
we apply the formulation presented for the simplest EMP molecule in the preceding sections 
to periodic structures of planar EMP crystals,
including a chain, ladder, and honeycomb network composed of $N$ domains.
To simplify the analysis,
we introduce the following vector notation for the two-component column matrix:
\begin{align}
 {\bf a}_i \equiv
 \begin{pmatrix}
  \tilde{a}_i \cr
  a_i
 \end{pmatrix}, \ \
 \tilde{\bf a}_i \equiv
 \begin{pmatrix}
  a_i \cr
  \tilde{a}_i
 \end{pmatrix}.
\end{align}
Note that a tilde rule is adopted,
namely the amplitude with a tilde is located in the first (second) component of ${\bf a}_i$ ($\tilde{\bf a}_i$).

\subsection{Chain}\label{sec:chain}

\begin{figure}[htbp]
 \begin{center}
  \includegraphics[scale=0.7]{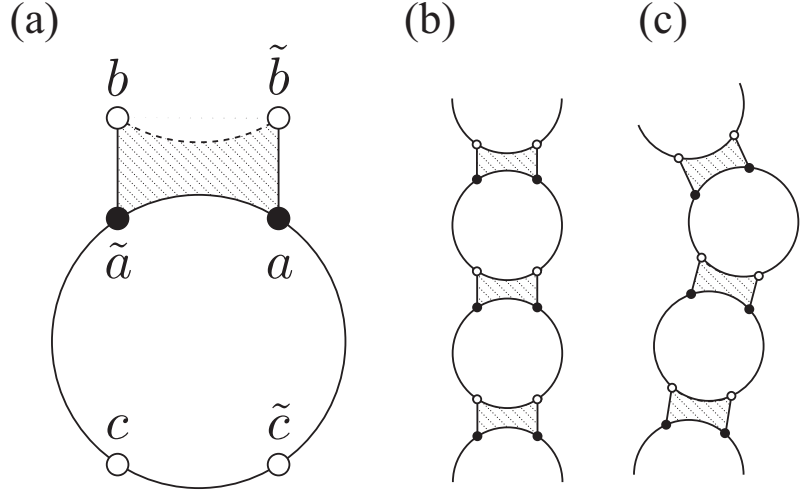}
 \end{center}
 \caption{{\bf Geometries of chain.}
 (a) The basic unit of a chain consists of a domain and coupled region.
 Note that the domain shape is arbitrary and that we assume it to be a circle here.
 The circumference of a domain is $L$, and the length of a coupled region is $\ell$.
 (b)
 When the circumferential distance on a domain between the two vertices $\tilde{c}$ and $a$ 
 is equal to that between $\tilde{a}$ and $c$ 
 ($\overline{\tilde{c} a}=\overline{\tilde{a} c}=\frac{L}{2}-\ell$), the chain is straight.
 Otherwise ($\overline{\tilde{c} a}=\frac{L}{2}-\ell-R_j$ and $\overline{\tilde{a} c}=\frac{L}{2}-\ell+R_j$
 for a $j$th domain), 
 the chain is deformed, as shown in (c).
 }
 \label{fig:block-A}
\end{figure}

A straight chain is formed when $N$ domains are aligned along a line. 
Figure~\ref{fig:block-A}(a) shows the constituents of the chain,
where the amplitudes of the vertices are related to each other by 
the boundary condition of the coupled region as 
$\tilde{\bf a} = T(\omega) {\bf b}$
and by a phase relationship of the uncoupled region as ${\bf c} = R_{\frac{L}{2}-\ell}(\omega)\tilde{\bf a}$.
Here, $R_{\frac{L}{2}-\ell}(\omega)$ originates from the phase accumulation caused by free propagation of EMPs
with a fixed chirality
from $\tilde{c}$ to $a$ and from $\tilde{a}$ to $c$ whose circular distance is $\frac{L}{2}-\ell$:
\begin{align}
 R_{\frac{L}{2}-\ell}(\omega)
 = 
 \begin{pmatrix}
  e^{-i\frac{\omega}{v} \left(\frac{L}{2}-\ell \right) } & 0 \cr
  0 & e^{+i\frac{\omega}{v} \left(\frac{L}{2}-\ell \right) } 
 \end{pmatrix}.
\end{align}
The elimination of $\tilde{\bf a}$ gives ${\bf c} = R_{\frac{L}{2}-\ell}(\omega)T(\omega){\bf b}$.
Because $\det(R_{\frac{L}{2}-\ell}(\omega)T(\omega))=1$ holds for any $\omega$,
we know from Bloch's theorem the existence of a unitary matrix $U$ and phase $\theta$ ($\in[0,\pi]$)
that satisfy
\begin{align}
 U R_{\frac{L}{2}-\ell}(\omega)T(\omega) U^\dagger = 
 \begin{pmatrix}
  e^{-i\theta} & 0 \cr
  0 & e^{+i\theta}
 \end{pmatrix}.
 \label{eq:URTU}
\end{align}
This is consistent with the characteristic equation 
$\lambda^2 - {\rm tr}(R_{\frac{L}{2}-\ell}(\omega)T(\omega)) \lambda + 1=0$,
and one may assume that the eigenvalues of $R_{\frac{L}{2}-\ell}(\omega)T(\omega)$ are $e^{\pm i \theta}$.
Equation~(\ref{eq:URTU}) leads to
the relation ${\rm tr}(R_{\frac{L}{2}-\ell}(\omega)T(\omega))=2 \cos \theta$, which is 
\begin{align}
 \cos\theta = \frac{1+r^2}{2r}
 \cos\left(\frac{\omega}{v} \left( \frac{L}{2}-\ell \right) \right)
 + \frac{1-r^2}{2r}
 \sin \left(\frac{\omega}{v} \left( \frac{L}{2}-\ell \right) \right)\cot\left( \frac{\omega \ell}{v_c} \right).
 \label{eq:A}
\end{align}
$\theta$ may be determined from Eq.~(\ref{eq:A})
as a function of $\omega$, which also specifies the dispersion relation of a chain.
Because the periodicity of $N$ domains is characterized by 
the boundary condition $\left[R(\omega) T(\omega)\right]^{N} = I$,
this condition discretizes $\theta$ 
through a constraint $N\theta = 2\pi n$, where $n$ is the wavenumber, and 
$\theta$ may be regarded as a continuum when $N\to \infty$.

\begin{figure}[htbp]
 \begin{center}
  \includegraphics[scale=0.7]{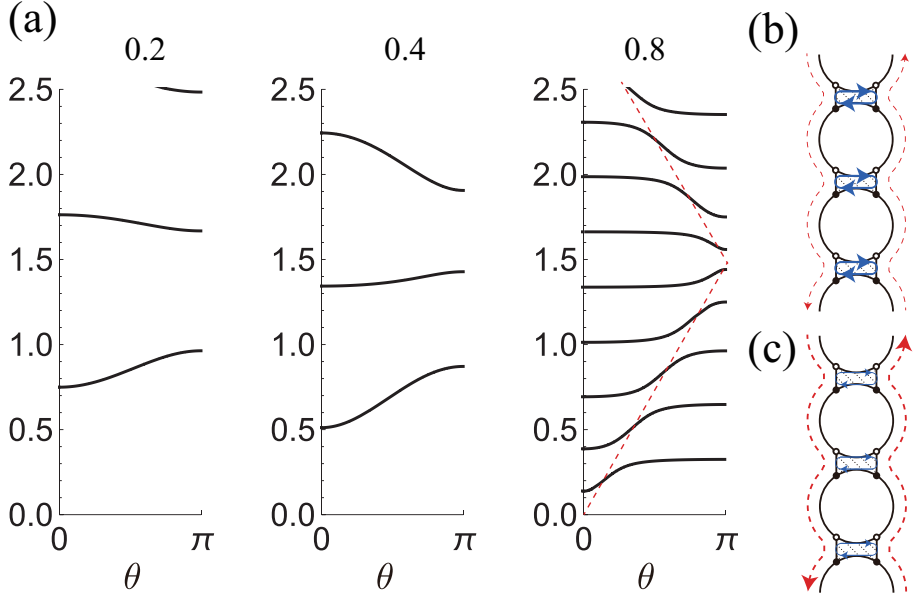}
 \end{center}
 \caption{{\bf Band structures of a chain.} (a)
 The dispersion is given as a function of $\theta \in [0,\pi]$ for different coupling strength $r=0.2$, $0.4$, and $0.8$.
 $\omega$ is normalized in units of $\frac{2\pi v}{L}$ with $L=6\ell$.
 Red dashed lines are the chiral dispersion of the original EMP of an isolated domain, which is the eigenmode 
 at the strong coupling limit.
 We note that There is an energy gap at $\theta=0$ which vanishes at the strong coupling limit as
 $\frac{\omega}{\frac{2\pi v}{L}} = \frac{1}{\pi} \sqrt{ \frac{(1-r)^2}{r r_c^2 (1-r_c)} }$.
 (b) and (c) illustrate the eigenmodes of a flat band and linear dispersion, respectively.
 }
 \label{fig:disp_chain}
\end{figure}

The band structure depends strongly on the coupling strength,
as shown in Fig.~\ref{fig:disp_chain}(a) for coupling constants 
$r=0.2$[left], $0.4$[middle] and $0.8$[right].
For a weak coupling ($r=0.2$),
a weak dispersive band appears near 
the fundamental excitation mode of a domain ($\omega = \frac{2\pi v}{L}$),
and energy gaps are formed between the subbands.
When $r=0.4$, the dispersive nature (or the bandwidth) is almost doubled.
There is a strong similarity between
the band structures shown in Fig.~\ref{fig:disp_chain}(a) 
and miniband structures calculated from the Kronig-Penny model
for periodic semiconductor superlattices.~\cite{Mendez1988,Bastard1982,Nakayama1996}
Indeed, as we will show in Appendix~\ref{app:KP}, 
the $T(\omega)$ matrix in Eq.~(\ref{eq:URTU})
can be constructed from physical variables of a binary superlattice.

In a strong coupling ($r=0.8$), the bandwidth of each subband is suppressed.
An overlap between the calculated dispersion and a linear dispersion of 
$\omega = \frac{v}{L/2-\ell}\theta$ 
[as expressed by red dashed lines in Fig.~\ref{fig:disp_chain}(a)] can be found at intervals.
Since $L/2-\ell$ is the effective unit-cell length along a chain, 
the linear dispersion may be expressed as $\omega=vk$, where $k$ is the wavevector along the chain
and Eq.~(\ref{eq:URTU}) shows that $\omega=vk$ becomes exact in the strong-coupling limit
because $T(\omega)$ becomes a unit matrix.
In fact, because the right-hand side of Eq.~(\ref{eq:A}) is singular at a multiple of $\omega_s$,
the subbands are separated by energy gaps formed at around $n \omega_s$.
In the gaps, $\theta$ is an imaginary number giving localized states.
The linear dispersion continuously changes into a flat band that represents the standing waves.
A flat band is mostly composed of the localized standing waves and 
is associated with a small component of a chiral wave in the uncoupled regions.
These dispersionless modes do not propagate along the chain.
A linear dispersion is mostly composed of the chiral wave in the uncoupled regions and 
is associated with a small component of the standing waves.
These dispersive modes propagate along the chain. 
The panels in Fig.~\ref{fig:disp_chain}(b) and (c) show these eigenmodes.

Since an ideal chain with a perfect periodicity does not exist in nature, 
we shall discuss a geometrical deformation of a chain.
When a straight chain is geometrically deformed locally by $R_j$, 
as shown in Fig.~\ref{fig:disp_chain}(c),
the matrix $R(\omega)$ acquires a $U(1)$ phase.
The periodic boundary condition is modified as
\begin{align}
 \prod_{j=1}^{N}
 \left[ e^{+i\frac{\omega}{v}R_j}
 \begin{pmatrix}
  e^{-i\frac{\omega}{v} (\frac{L}{2}-\ell)} & 0 \cr
  0 & e^{+i\frac{\omega}{v} (\frac{L}{2}-\ell)} 
 \end{pmatrix}
 T(\omega)
 \right] = I.
\end{align}
When $R\equiv \sum_{j=1}^N R_j=0$, the effect of the local deformation is removed, 
which is similar to the pure gauge degree of freedom in gauge theories.
When $R \ne 0$, a chain is not a straight line but a closed curve. 
Such a change in global topology does not alter the band structure but 
may cause a physical effect, namely a shift in the wavenumber 
\begin{align}
 \theta = \frac{2\pi}{N} \left( n 
 - \frac{\omega R}{2\pi v}\right).
 \label{eq:theta}
\end{align}
We apply this result to understand the effect of a geometrical change from a straight line to a square.
Suppose eight domains $(N=8)$ with $L=8\ell$ are aligned to form a straight line. 
It can be deformed into a square by setting $R=8\ell$.
In the weak coupling limit, we may assume $\omega = \frac{2\pi v}{L}$.
By putting it into Eq.~(\ref{eq:theta}), we have $\theta = \frac{2\pi}{N}(n-1)$.
However, a shift in $\theta$ would be difficult to validate experimentally because 
a planar periodic crystal must be modified to obtain an output signal.
A $U(1)$ phase can be irrelevant to physical observables like the reflectance or transmittance,
because they are given by the amplitude absolute square.

The results for the two coupled domains, such as in Fig.~\ref{fig:2link}(a) in Sec.~\ref{sec:twodomains},
are approximately embedded into the band structure in Fig.~\ref{fig:disp_chain}(a)
at $\theta=0$ and $\pi$.
Specifically, for the weak coupling, we may rewrite Eq.~(\ref{eq:ab_2rings}) as 
$a/b=- \cos\theta$ by using Eq.~(\ref{eq:A}).
This is shown by replacing $L$ in Eq.~(\ref{eq:ab_2rings}) with $\frac{L}{2}$,
and the remaining $\frac{L}{2}$ is used to obtain the minus sign in $a/b=-\cos\theta$, 
where the phase relationship between $a$ and $b$ is reversed for the case that $\omega \simeq \frac{2\pi v}{L}$
because $\tan(x+\pi)=\tan(x)$ and $\cos(x+\pi)=-\cos(x)$.

\subsection{Ladder}

\begin{figure}[htbp]
 \begin{center}
  \includegraphics[scale=0.7]{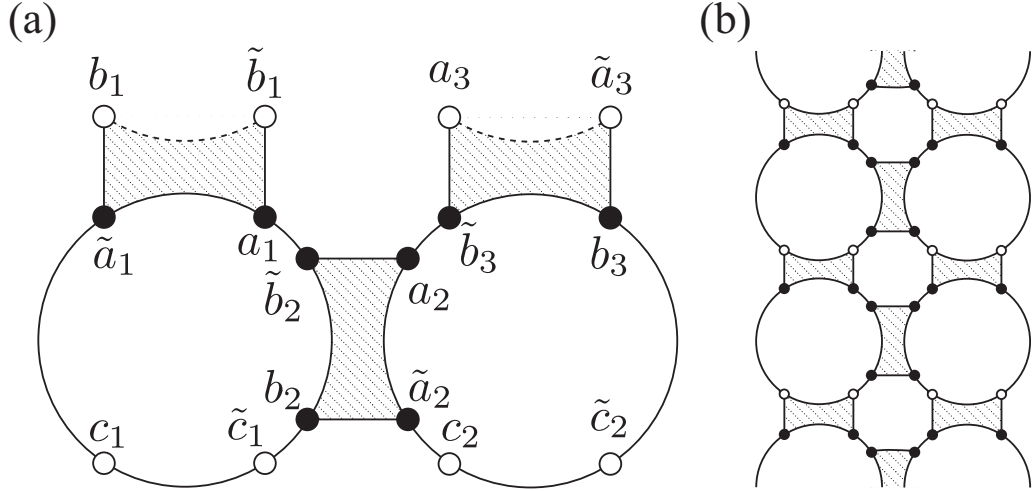}
 \end{center}
 \caption{{\bf Geometry of ladder.} (a)
 The basic unit of a ladder consists of the two domains and three coupled regions.
 The circumference of a domain is $L$, and the length of a coupled region is $\ell$.
 The circumferential distance between $a_1$ and $\tilde{b}_2$ 
 ($\overline{a_1 \tilde{b}_2}$)
 is equal to $\overline{a_2 \tilde{b}_3}$, $\overline{b_2 \tilde{c}_1}$, and $\overline{\tilde{a}_2 c_2}$, 
 which is given by $s\equiv \frac{L}{4}-\ell$.
 (b) The domain structure of a straight ladder. 
 }
 \label{fig:block-AB}
\end{figure}

We obtain a straight ladder by interconnecting the two basic units of a chain as shown in Fig.~\ref{fig:block-AB}(a)
and by identifying ${\bf c}_2$ and ${\bf c}_1$ with ${\bf a}_3$ and ${\bf b}_1$, respectively,
as shown in Fig.~\ref{fig:block-AB}(b).
Due to the chirality,
a ladder includes two input channels (say $\tilde{c}_1$ and $\tilde{c}_2$).
Thus, 
a ladder corresponds to a device that can reflect or transmit the two wave signals.

To calculate its band structure, 
we need to construct a $4\times 4$ matrix that satisfies
\begin{align}
 \begin{pmatrix}
  {\bf c}_2 \cr {\bf c}_1
 \end{pmatrix}
 = M(\omega)
 \begin{pmatrix}
  {\bf a}_3 \cr {\bf b}_1
 \end{pmatrix}.
\end{align}
In addition to the $T$ matrix satisfying 
$\tilde{\bf b}_3=T(\omega) {\bf a}_3$ and $\tilde{\bf a}_1=T(\omega) {\bf b}_1$,
let us introduce a $2\times 2$ matrix for the interconnected region between the two domains.
\begin{align}
 \begin{pmatrix}
  \tilde{a}_2 \cr b_2
 \end{pmatrix}
 =S(\omega)
 \begin{pmatrix}
  a_2 \cr \tilde{b}_2
 \end{pmatrix}.
\end{align}
This $S(\omega)$ matrix is known from the boundary condition Eq.~(\ref{eq:bc}) as
\begin{align}
 S(\omega) \equiv
 U R_c(\omega) U^{-1} = \cos\left(\frac{\omega \ell}{v_c}\right)I+i\sin\left(\frac{\omega \ell}{v_c}\right) W,
\end{align}
where
\begin{align}
  U \equiv 
 \begin{pmatrix}
  1 & r \cr
  r & 1
 \end{pmatrix}, \ \
 R_c(\omega)
 \equiv 
 \begin{pmatrix}
  e^{+i\frac{\omega \ell}{v_c}} & 0 \cr
  0 & e^{-i\frac{\omega \ell}{v_c}}
 \end{pmatrix}, \ \
 W = \frac{1+r^2}{1-r^2}
 \begin{pmatrix}
  1 & -\frac{2r}{1+r^2} \cr
  \frac{2r}{1+r^2} & -1
 \end{pmatrix}.
\end{align}
We note that because $W^2 = I$,
$S(\omega) = e^{i \left(\frac{\omega \ell}{v_c}\right) W}$.
This expression may be used to simplify some calculation.

\begin{figure}[htbp]
 \begin{center}
  \includegraphics[scale=0.5]{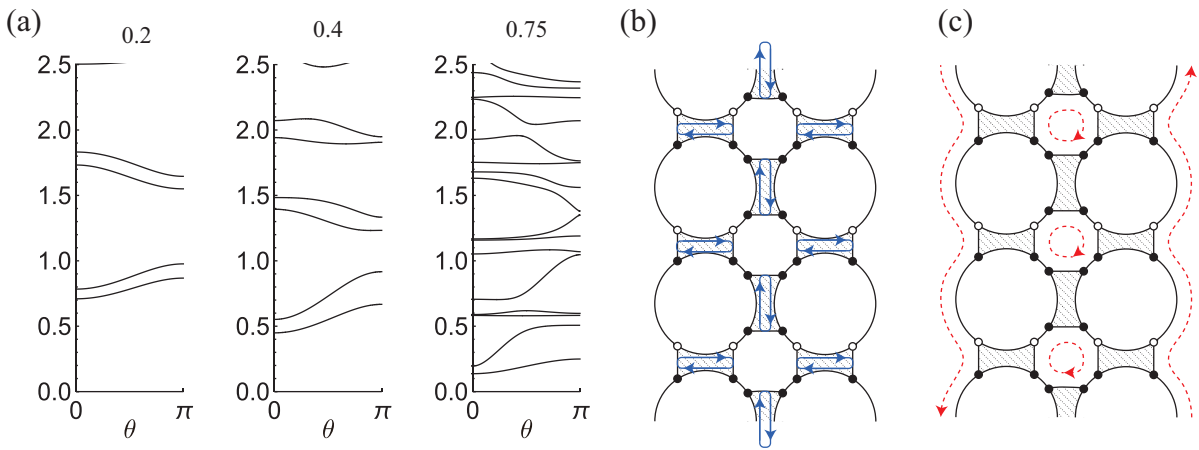}
 \end{center}
 \caption{{\bf Band structures of a ladder.} (a)
 Dispersion is plotted as a function of $\theta \in [0,\pi]$.
 $\omega$ is in units of $\frac{2\pi v}{L}$, where $L=8\ell$ is assumed.
 (b) and (c) Eigenmodes of a flat band and linear dispersion, respectively.
 }
 \label{fig:disp_ladder}
\end{figure}

For the free propagation of the EMP in the uncoupled regions, we have
\begin{align}
 \begin{pmatrix}
  a_2 \cr \tilde{b}_2
 \end{pmatrix}
 =R_s(\omega)
 \begin{pmatrix}
  \tilde{b}_3 \cr a_1
 \end{pmatrix}, \ \ 
 R_s(\omega)
 \equiv 
 \begin{pmatrix}
  e^{+i\frac{\omega s}{v}} & 0 \cr
  0 & e^{-i\frac{\omega s}{v}}
 \end{pmatrix}.
\end{align}
Therefore, we obtain 
\begin{align}
 &
 \begin{pmatrix}
  c_2 \cr \tilde{c}_1
 \end{pmatrix}
 = R_s(\omega) S(\omega)R_s(\omega)
 \begin{pmatrix}
  \tilde{b}_3 \cr a_1
 \end{pmatrix},
 \\
 &
 \begin{pmatrix}
  \tilde{c}_2 \cr c_1
 \end{pmatrix}
 = 
 \begin{pmatrix}
  e^{-i\frac{\omega}{v} (\frac{L}{2}-\ell)} & 0 \cr
  0 & e^{+i\frac{\omega}{v} (\frac{L}{2}-\ell)}
 \end{pmatrix}
 \begin{pmatrix}
  b_3 \cr \tilde{a}_1
 \end{pmatrix}.
\end{align}
Finally, the explicit form of the $4\times 4$ matrix $M(\omega)$ is given by
\begin{align}
 M(\omega) =
 \begin{pmatrix}
  e^{-i\frac{\omega}{v} (\frac{L}{2}-\ell)} & 0 & 0 \cr
  0 & R_s(\omega) S(\omega) R_s(\omega) & 0 \cr 
  0 & 0 & e^{+i\frac{\omega}{v} (\frac{L}{2}-\ell)} 
 \end{pmatrix}
 \begin{pmatrix}
  T(\omega) & 0 \cr
  0 & T(\omega)
 \end{pmatrix}.
 \label{eq:ladderM}
\end{align}
The characteristic equation of $M(\omega)$ is 
written as a symmetric form 
$\lambda^4 + A(\omega) \lambda^3 + B(\omega) \lambda^2 + A(\omega) \lambda + 1=0$,
with functions $A(\omega)=-{\rm tr}(M(\omega))$ and $B(\omega)=\frac{1}{2}(A(\omega)^2-{\rm tr}(M(\omega)^2))$.
By setting $\lambda = e^{i\theta}$, we rewrite this as 
\begin{align}
 \left( \cos \theta + \frac{A(\omega)}{4} \right)^2 = \frac{8+A(\omega)^2-4B(\omega)}{16}.
\end{align}
By solving it with respect to $\theta$,
we plot the dispersion relation in Fig.~\ref{fig:disp_ladder}(a)
for coupling constants ($r=0.2$ [left], $0.4$ [middle], and $0.75$ [right]).
For a weak coupling ($r=0.2$),
the dispersion appears near $\omega = \frac{2\pi v}{L}$, 
which is the fundamental excitation mode of a domain.
According to the two domains in the unit cell of a ladder, 
two dispersion curves appear as a pair in the weak coupling.
For a strong coupling ($r=0.75$), 
the standing wave modes ($\omega_s$) appear as flat bands between 0.5 and 0.6.
These modes are also localizing at an interconnected region between the two domains of a unit cell.
They are nearly degenerate because $S(n\omega_s)=(-1)^n I$ holds and therefore 
Eq.~(\ref{eq:ladderM}) consists of the same $2\times 2$ matrix in a diagonal form.

In the strong coupling limit, since $T(\omega) \to 1$,
we can expect that the possible modes of the system are divided into a counter propagating 
(outer) edge modes and other inner modes.
The latter $-$EMPs rotating around each hole of the system$-$ have a higher energy 
$\omega= \frac{\pi v}{2s}$, which is visible as an almost flat band.

\subsection{Honeycomb}

\begin{figure}[htbp]
 \begin{center}
  \includegraphics[scale=0.5]{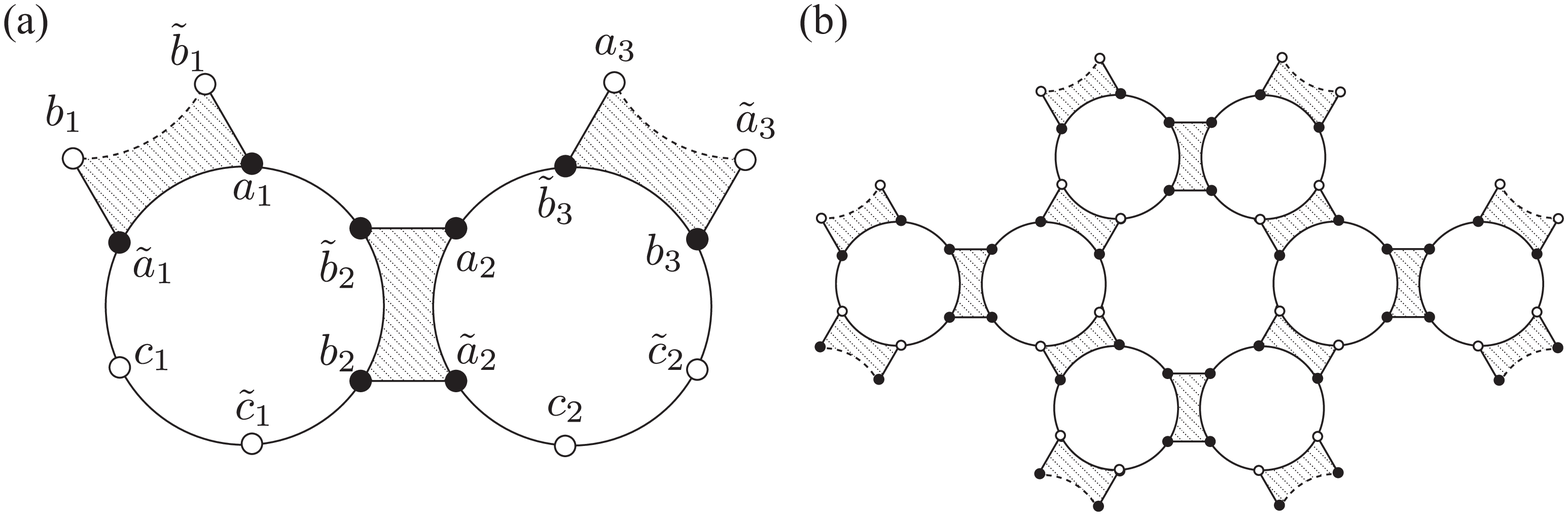}
 \end{center}
 \caption{{\bf Geometry of honeycomb lattice.} (a) and (b)
 The basic unit of a honeycomb network is given by changing the vertex positions of the basic unit of a ladder. 
 }
 \label{fig:block-ABC}
\end{figure}

As shown in Figs.~\ref{fig:block-ABC}(a) and (b),
a honeycomb network can be obtained by slightly modifying the basic unit of a ladder.
The circumferential distance between all nearest neighbor vertices in the uncoupled regions 
must be the same;
the circumferential distance $\overline{a_1 \tilde{b}_2}$
is equal to $\overline{c_1 \tilde{a}_1}$, $\overline{b_2 \tilde{c}_1}$,
$\overline{a_2 \tilde{b}_3}$, $\overline{b_3 \tilde{c}_2}$, and $\overline{c_2 \tilde{a}_2}$.

The corresponding matrix is given by
replacing $\frac{L}{2}-\ell$ of $e^{\pm i \frac{\omega}{v}(\frac{L}{2}-\ell)}$ in Eq.~(\ref{eq:ladderM})
with $s$, where $s=\frac{L}{3}-\ell$, as
\begin{align}
 M(\omega) =
 \begin{pmatrix}
  e^{-i\frac{\omega s}{v}} & 0 & 0 \cr
  0 & R_s(\omega)S(\omega)R_s(\omega) & 0 \cr 
  0 & 0 & e^{+i\frac{\omega s}{v}} 
 \end{pmatrix}
 \begin{pmatrix}
  T(\omega) & 0 \cr
  0 & T(\omega)
 \end{pmatrix}.
\end{align}
Two adjacent units can be connected by a twisted boundary condition:
\begin{align}
 \begin{pmatrix}
  {\bf c}_2 \cr {\bf c}_1
 \end{pmatrix}
 =
 \begin{pmatrix}
  0 & e^{i\theta_1} I \cr
  e^{i\theta_2} I & 0 
 \end{pmatrix}
 \begin{pmatrix}
  {\bf a}_3 \cr {\bf b}_1
 \end{pmatrix},
\end{align}
where $\theta_1 \equiv \theta +\phi$ and $\theta_2 \equiv \theta -\phi$.
Therefore, we need to diagonalize the following $4\times 4$ matrix.
\begin{align}
 M_{h}(\omega;\phi) = 
 \begin{pmatrix}
  0 & e^{+i\phi} \cr
  e^{-i\phi} & 0 
 \end{pmatrix}
 M(\omega).
\end{align}
The characteristic equation of $M_{h}$ is 
written as $\lambda^4 + 2 \lambda^3 \cos\phi  + \tilde{B}(\omega;\phi) \lambda^2 + 2 \lambda \cos\phi + 1=0$ 
where $\tilde{B}(\omega;\phi)=2\cos^2(\phi)-\frac{1}{2}{\rm tr}(M_h(\omega;\phi)^2)$.
Setting $\lambda = e^{i\theta}$ leads to 
\begin{align}
 \left( \cos \theta + \frac{\cos\phi}{2} \right)^2 = \frac{\cos^2\phi + 2-\tilde{B}(\omega;\phi)}{4}.
\end{align}
By solving it with respect to $\theta$ with $\phi=0$,
we can obtain the dispersion relation along $\Gamma{\rm K}$.

\begin{figure}[htbp]
 \begin{center}
  \includegraphics[scale=0.5]{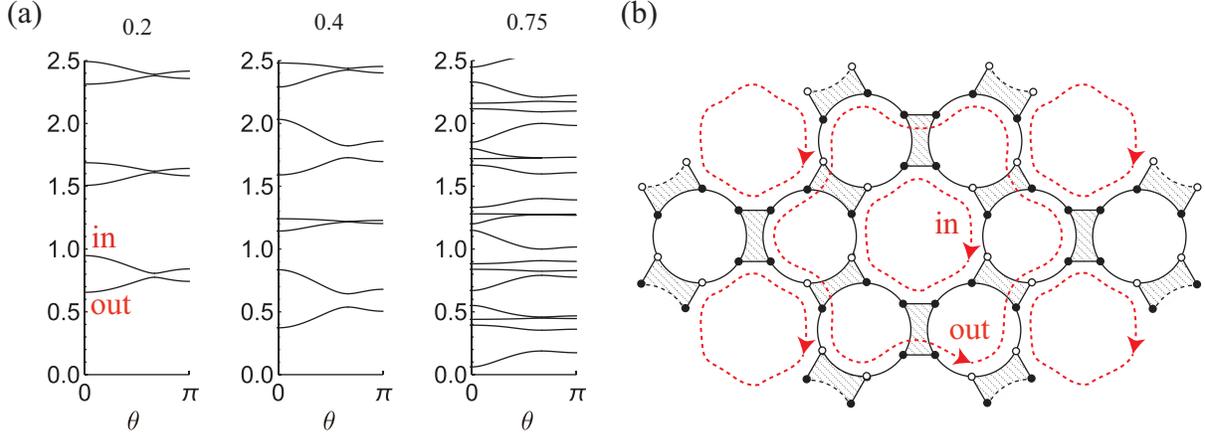}
 \end{center}
 \caption{{\bf Band structure of a honeycomb lattice.} (a)
 Dispersion (along $\Gamma{\rm K}$) is shown as a function of $\theta \in [0,\pi]$.
 $\omega$ is in units of $\frac{2\pi v}{L}$, where $L=6\ell$.
 (b) Two approximate eigenmodes of $\theta=0$ are illustrated by the dashed arrows.
 }
 \label{fig:disp_hc}
\end{figure}

Figure~\ref{fig:disp_hc}(a) shows the band structures for coupling constants 
($r=0.2$ [left], $0.4$ [middle], and $0.75$ [right]).
For the weak coupling ($r=0.2$),
the energy band has a small energy gap at the K point ($\theta=\frac{2\pi}{3}$)
for the lowest two energy subbands.
It is difficult to see due to the resolution, but a small gap opens for the higher subbands.
The gap of the fundamental subbands increases with increasing coupling strength.

We discuss the result using two possible modes of the system.
One is the mode rotating around each hexagonal hole 
(or the inner edge of a hexagonal ring) [see Fig.~\ref{fig:disp_hc}(b)].
This mode has energy similar to that of the fundamental mode $\omega= \frac{2\pi v}{L}$ 
and appears as the second subband at $\theta=0$ in the weak coupling regime.
The other is the mode rotating around the outer edge of a hexagonal ring, 
which has a larger perimeter than the mode rotating around a hexagonal hole.
This appears as the lowest energy subband at $\theta =0$ in the weak coupling regime.
These two modes are coupled together to form real eigenmodes.
In the strong coupling, the flat-band nature of the subband
with the second lowest energy is noticed. 
The standing waves of the coupled regions are weakly interacting with each other 
and form a nearly flat band.

The energy band structure of honeycomb networks is not as well understood as it is for the chain.
For example, the energy positions of the Dirac cone are not identified as a function of $\omega_s$.
The clarification of such a problem requires more study.
Note also that our honeycomb network differs greatly from a chain and ladder 
in the sense that it does not have an outer boundary.
It is not evident for the general coupling strength 
whether a finite honeycomb network can support EMPs at the periphery.
Introducing an outer edge to the honeycomb network would require some additional effort, 
which is beyond the scope of this paper.
The matrix formulation we have developed for the EMP molecule and crystals 
is amenable to a transfer matrix method, with which we can calculate physical observables
of finite periodic systems composed of $N$ domains, which we will show in detail in a subsequent paper.

\section{Discussion}

Strong coupling is intriguing 
from various points of view, including a perfect transmission mode and flat band.
Graphene has the advantage of realizing a strong coupling.
Brasseur {\it et al.} achieved $r$ as large as 0.55 for two domains separated by 
a narrow etched line (0.3 $\mu$m width) in graphene.~\cite{Brasseur2017} 
This should be compared with $r\sim 0.04$ 
obtained for the edge channels defined by a metal gate (1 $\mu$m width) in a GaAs/AlGaAs heterostructure.~\cite{Kamata2014}
The $r$ values differ partly because the inter-edge capacitive coupling is suppressed by the 
screening effect of the metal gate
and because the sharp edge potential of graphene prevents formation of the depletion layer
(which increases virtually the width).

When two domains are positioned very close to each other for strong coupling, 
the validity of the description on the coupled region using a large coupling strength is not evident.
Suppose that two domains merge into a single domain.
The coupled region becomes the bulk region, where an EMP does not exist.
The absence of low-energy excitation in the bulk is in sharp contrast to 
the result that many states are condensed into zero-energy in the strong coupling limit.
Therefore, there may be a breakdown in describing the coupled region 
with a very narrow inter-domain distance in terms of a large $r$.
We are speculating that this problem is fundamentally related to the inter-domain charge transfer 
caused by electron tunneling.

Our description of an EMP crystal in this manuscript looks unrelated to quantum mechanics;
however, an essential feature of quantum theory is partly built-in.
Suppose that for an EMP molecule, 
an EMP pulse in the first domain enters the coupled region. 
In the second domain, at the boundary $x=0$, a pair creation from the vacuum takes place. 
This is a process of the creation of a particle and antiparticle,
which is a phenomenon handled by the quantum field theory.
We also note that for the diatomic EMP molecule discussed in Sec.~\ref{sec:twodomains},
the energy density may be identified as a potential energy:
\begin{align}
 H(x) = 
 \begin{pmatrix}
  \rho_1(x) & \rho_2(x)
 \end{pmatrix}
 \begin{pmatrix}
  V_1(x) \cr V_2(x)
 \end{pmatrix}=
 \frac{1}{c_{ch}}
 \begin{pmatrix}
  \rho_1(x) & \rho_2(x)
 \end{pmatrix}
 \begin{pmatrix}
  1-\delta & \delta \cr \delta & 1 - \delta
 \end{pmatrix}
 \begin{pmatrix}
  \rho_1(x) \cr \rho_2(x)
 \end{pmatrix}.
\end{align}
By Eq.~(\ref{eq:vrho}),
$H(x)$ is rewritten as a quadratic form in the charge density variables 
$\rho_1$ and $\rho_2$ (or current densities $j_1$ and $j_2$).
This is consistent with a quantum mechanical Hamiltonian density, 
by which a quantum mechanical description of the system is possible based on the U(1) current algebra.~\cite{WEN1992}

There are some possible extensions of the work described in this paper. 
One is to include the spin (current).
For the QHE with $\nu =2$, (dynamical) 
charge and spin currents coexist at the edge of a single domain.
Though it is not evident that the formulation based on a capacitive interaction (between different domains)
holds for this case (of different edge channels in the same domain), 
recent experiments show that this is indeed valid.~\cite{Hashisaka2017}
It would also be interesting to include the opposite chirality in the same domain, 
which is expected for a quantum spin Hall effect.
From a theoretical point of view, if the spin degrees of freedom is replaced with pseudospin,
such edge plasmon crystal without an external magnetic field is relevant to 
the plasmons observed in doped carbon nanotubes 
(albeit with a difference in spatial scales).~\cite{Uryu2018,Satco2019,Sasaki2018,Yanagi2018}
Though this appears to be an impossible geometry for EMPs,
azimuthal plasmons in doped carbon nanotubes (CNTs) 
can be treated as a circular current in two dimensions,
if the domain is regarded as the cross section of a CNT.
This is an issue to which the results of this paper could be applied.
We speculate that some discrepancy between theory and experiments found recently~\cite{Sasaki2020}
may be partly resolved by a capacitive coupling between the plasmons.

\section{Summary}

The band structures of EMP crystals (chain, ladder, and honeycomb network) 
were calculated based on the continuity of the current density with a transfer matrix method.
The calculated results are explained by the eigen modes of an EMP molecule composed of two equivalent atoms (domains).
We have discussed the effect of a geometrical deformation of a chain 
on the wavenumber in terms of a gauge degree of freedom.
We pointed out an interesting similarity between EMP crystals and layered materials (superlattices).

\section*{Acknowledgments}

The author thanks M. Hashisaka and K. Muraki for proposing the problem. 
The author is also grateful to N. Kumada for his outstanding instruction.

\appendix

\section{Domains with opposite magnetic field directions}\label{app:sd}

\begin{figure}[htbp]
 \begin{center}
  \includegraphics[scale=0.8]{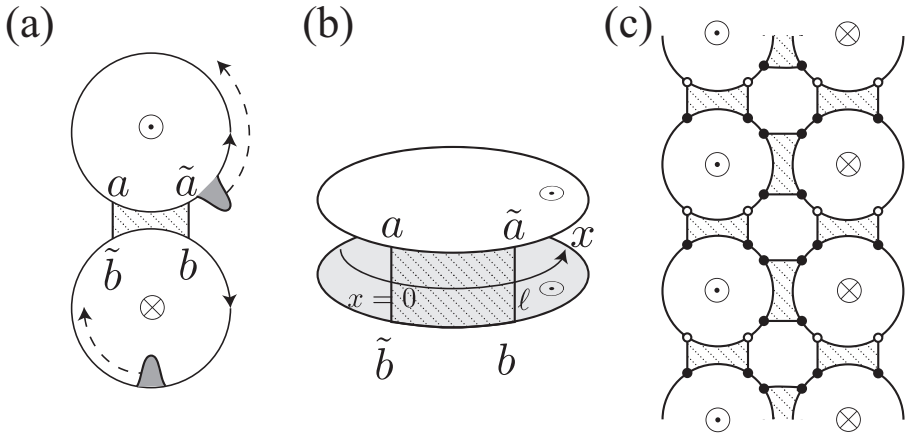}
 \end{center}
 \caption{{\bf Geometries of two domains with a staggered magnetic field.}
 The geometry in two dimensions (a) corresponds to (b) in three dimensions.
 An example of a planar periodic crystals is shown in (c).
 }
 \label{fig:emp_stagger}
\end{figure}

We show a planar geometry 
composed of two capacitively coupled domains having opposite magnetic field directions
in Fig.~\ref{fig:emp_stagger}(a).
This configuration of the staggered magnetic field appears to be a little unrealistic.
However, as shown in Fig.~\ref{fig:emp_stagger}(b),
the topologically equivalent configuration in three-dimensions corresponds to a uniform magnetic field,
as opposed to that in Fig.~\ref{fig:emp_coupled}(c),
and thus turns out to be a more realistic.
Indeed, when the two domains are merged into a single domain in Fig.~\ref{fig:emp_stagger}(b) by setting 
the inter-domain distance to zero and also $\ell \to L$, 
this serves as a model for co-propagating spin-polarized edge channels in 
a single domain with $\nu=2$ QHE.~\cite{Hashisaka2017,Hashisaka2018a}
The situation is also relevant to a capacitor in an external magnetic field,
for which the following analysis would have direct relevance.

The study of the two domains is rather straightforward. 
The unique modification that we need to apply is 
\begin{align}
 \begin{pmatrix}
  j_1(x) \cr
  j_2(x)
 \end{pmatrix}
 = \sigma_{xy}
 \begin{pmatrix}
  1 & 0 \cr 
  0 & 1
 \end{pmatrix}
 \begin{pmatrix}
  V_1(x) \cr V_2(x)
 \end{pmatrix},
\end{align}
instead of Eq.~(\ref{eq:jh}).
By repeating the same analysis,
we have for the coupled region $x \in [0,\ell]$ two modes with the same chirality.
The current density is written as
\begin{align}
 \begin{pmatrix}
  j_1(x) \cr
  j_2(x)
 \end{pmatrix}
 = 
 \alpha_R 
 \begin{pmatrix}
  1 \cr 1 
 \end{pmatrix}
 e^{+i\frac{\omega}{v} x}
 +
 \beta_R
 \begin{pmatrix}
  1 \cr -1
 \end{pmatrix}
 e^{+i\frac{\omega}{v_s} x}.
\end{align}
The velocities of the two modes are $v$ and $v_s \equiv (1-2\delta)v$.
Note that the renormalized velocity $v_s$ differs from $v_c$ 
by the multiplicative factor of $\sqrt{1-2\delta}$.
Due to the continuity condition of the current density, 
the current amplitudes at the vertices are related by 
\begin{align}
 \begin{pmatrix}
  \tilde{a} \cr b 
 \end{pmatrix}
 = V(\omega)
 \begin{pmatrix}
  a \cr \tilde{b}
 \end{pmatrix},
 \label{eq:Vma}
\end{align}
where the matrix $V$ is defined as
\begin{align}
 V(\omega) \equiv 
 \begin{pmatrix}
  1 & 1 \cr 1 & -1
 \end{pmatrix}
 \begin{pmatrix}
  e^{i\frac{\omega \ell}{v}} & 0 \cr 0 & e^{i\frac{\omega \ell}{v_s}}
 \end{pmatrix}
 \begin{pmatrix}
  1 & 1 \cr 1 & -1
 \end{pmatrix}^{-1}.
\end{align}
Applying the phase relations $a=e^{i\frac{\omega(L-\ell)}{v}}\tilde{a}$
and $\tilde{b}=e^{i\frac{\omega(L-\ell)}{v}}b$ to Eq.~(\ref{eq:Vma}),
we obtain the frequency of a non-bonding state as $\omega = \frac{2\pi v}{L}n$
and that of a bonding orbital as 
\begin{align}
 \omega = \frac{2\pi n}{\frac{L-\ell}{v}+\frac{\ell}{v_s}}.
\end{align}
For the special case of $\ell=L$, $\omega = \frac{2\pi v_s}{L}n$.
More generally, 
in order to make the coupled region a limited part of the domain, 
it is necessary to prepare two domains with different diameters, but such details are ignored here.
The $V$ matrix is used to calculate the band structure of a ladder shown in Fig.~\ref{fig:emp_stagger}(c), 
which can be obtained by diagonalizing the matrix $M$ given by
\begin{align}
 M(\omega) =
 \begin{pmatrix}
  e^{-i\frac{\omega}{v} (\frac{L}{2}-\ell)} & 0 & 0 \cr
  0 & e^{i\frac{\omega s}{v}} V(\omega) e^{i\frac{\omega s}{v}} & 0 \cr 
  0 & 0 & e^{-i\frac{\omega}{v} (\frac{L}{2}-\ell)} 
 \end{pmatrix}
 \begin{pmatrix}
  T(\omega) & 0 \cr
  0 & T^*(\omega)
 \end{pmatrix}.
\end{align}
It is also useful to define the matrix 
\begin{align}
 T_s(\omega) \equiv 
 \begin{pmatrix}
  1 & 1 \cr e^{i\frac{\omega \ell}{v}} & e^{i\frac{\omega \ell}{v_s}}
 \end{pmatrix}
 \begin{pmatrix}
  1 & 0 \cr 0 & -1
 \end{pmatrix}
 \begin{pmatrix}
  1 & 1 \cr e^{i\frac{\omega \ell}{v}} & e^{i\frac{\omega \ell}{v_s}}
 \end{pmatrix}^{-1}
\end{align}
that relates the current density of one domain to that of the other domain as
\begin{align}
 \begin{pmatrix}
  a \cr \tilde{a}
 \end{pmatrix}
 = T_s(\omega)
 \begin{pmatrix}
  \tilde{b} \cr b
 \end{pmatrix}.
 \label{eq:Ts}
\end{align}
The $T_s$ matrix satisfies $T_s^{-1}=T_s$ and $\det(T_s)=-1$.

\section{Correspondence to Kronig-Penny model}\label{app:KP}

The Kronig-Penney model is a model for an electron in a one-dimensional periodic potential.~\cite{Kronig1931}
In this appendix, we show that 
an EMP chain bears a remarkable similarity to the electron in a superlattice
by studying the model using the method developed for EMPs.

Suppose that the unit cell of the superlattice consists of two layers (A and B) with a potential difference $V$
due to band discontinuity.
In the unit cell from $x=-b$ to $a$, 
the wave function is written as a sum of left and right moving waves as
\begin{align}
 \psi(x) = 
 \begin{cases}
  & A e^{+ik_A x} + \tilde{A} e^{-ik_A x} \ \ (0 < x < a) \\
  & B e^{+ik_B x} + \tilde{B} e^{-ik_B x} \ \ (-b < x < 0)
 \end{cases}
\end{align}
where the wavevector $k_A$ and $k_B$ are related to the energy eigenvalue $E$ 
by the Schr\"{o}dinger equation for the nonrelativistic electron with the effective mass $m$,
\begin{align}
 \frac{\hbar^2}{2m} k_A^2 = E, \ \ 
 \frac{\hbar^2}{2m} k_B^2 + V = E.
 \label{eq:Sch}
\end{align}
Since the wave function and its first derivative with respect to $x$ 
must be continuous at the boundary between layers A and B, we obtain
the boundary condition at $x=0$:
\begin{align}
 \begin{pmatrix}
  1 & 1 \cr
  k_A & -k_A
 \end{pmatrix}
 \begin{pmatrix}
  A \cr \tilde{A}
 \end{pmatrix}
 =
 \begin{pmatrix}
  1 & 1 \cr
  k_B & -k_B
 \end{pmatrix}
 \begin{pmatrix}
  B \cr \tilde{B}
 \end{pmatrix}.
\end{align}
From this, we define two matrices $T_{AB}$ and $T_{BA}$ as follows:
\begin{align}
 T_{AB}
 \equiv
 \begin{pmatrix}
  1 & 1 \cr
  k_A & -k_A
 \end{pmatrix}^{-1}
 \begin{pmatrix}
  1 & 1 \cr
  k_B & -k_B
 \end{pmatrix},
 \ \
 T_{BA} \equiv T_{AB}^{-1}
\end{align}
so that 
\begin{align}
 \begin{pmatrix}
  A \cr \tilde{A}
 \end{pmatrix}
 =T_{AB}
 \begin{pmatrix}
  B \cr \tilde{B}
 \end{pmatrix}, 
 \ \
 \begin{pmatrix}
  B \cr \tilde{B}
 \end{pmatrix}
 =T_{BA}
 \begin{pmatrix}
  A \cr \tilde{A}
 \end{pmatrix}.
\end{align}
Using these equations, we construct the transfer matrix that 
relates the wave function at layer A to that at the nearest neighbor of layer A as~\cite{Bastard1982}
\begin{align}
 \begin{pmatrix}
  A \cr \tilde{A}
 \end{pmatrix}_{j+1}
 =
 T_{AB}
 \begin{pmatrix}
  e^{+ik_B b} & 0 \cr
  0 & e^{-ik_B b} 
 \end{pmatrix}
 T_{BA}
 \begin{pmatrix}
  e^{+ik_A a} & 0 \cr
  0 & e^{-ik_A a} 
 \end{pmatrix}
 \begin{pmatrix}
  A \cr \tilde{A}
 \end{pmatrix}_{j}.
\end{align}
Finally, applying Bloch's theorem to the diagonalized transfer matrix, we obtain
\begin{align}
 U \left[ 
 T_{AB}
 \begin{pmatrix}
  e^{+ik_B b} & 0 \cr
  0 & e^{-ik_B b} 
 \end{pmatrix}
 T_{BA}
 \begin{pmatrix}
  e^{+ik_A a} & 0 \cr
  0 & e^{-ik_A a} 
 \end{pmatrix}
 \right] U^\dagger = 
 \begin{pmatrix}
  e^{-ik(a+b)} & 0 \cr
  0 & e^{+ik(a+b)}
 \end{pmatrix},
 \label{eq:UTRTRU}
\end{align}
which leads to 
\begin{align}
 {\rm tr}\left[ 
 T_{AB}
 \begin{pmatrix}
  e^{+ik_B b} & 0 \cr
  0 & e^{-ik_B b} 
 \end{pmatrix}
 T_{BA}
 \begin{pmatrix}
  e^{+ik_A a} & 0 \cr
  0 & e^{-ik_A a} 
 \end{pmatrix}
 \right] = 2\cos[k(a+b)].
\end{align}
This may be rewritten as the compact form
\begin{align}
 \cos[k(a+b)]
 = \cos(k_A a)\cos(k_B b)-
 \frac{1}{2} \left(\frac{k_A}{k_B} + \frac{k_B}{k_A} \right) \sin(k_A a)\sin(k_B b).
 \label{eq:KPR}
\end{align}
By putting Eq.~(\ref{eq:Sch}) into Eq.~(\ref{eq:KPR}), 
the possible energies that the electron can occupy (or miniband structures $E_n(k)$) are obtained
as a function of the wavevector $k$.

The mathematical similarity between Eq.~(\ref{eq:UTRTRU}) and Eq.~(\ref{eq:URTU}) becomes more evident
for the localized states at layer B 
with $k_B=i/\xi_B$, where $\xi^{-1}_B=\sqrt{2m(V-E)}/\hbar$ ($0 \le E \le V$) is the inverse of the decay length.
This stems from the fact that the $T(\omega)$ matrix of Eq.~(\ref{eq:URTU}) [or Eq.~(\ref{eq:Tmat})]
can be reproduced from 
\begin{align}
 T_{AB}
 \begin{pmatrix}
  e^{+ik_B b} & 0 \cr
  0 & e^{-ik_B b} 
 \end{pmatrix}
 T_{BA}
 \label{eq:TABTBA}
\end{align}
by putting $k_B =i/\xi_B$ into it as
\begin{align}
 \frac{1}{-2i r \sin 2\phi}
 \begin{pmatrix}
  e^{-2i\phi}-r^2 e^{+2i\phi} & -(1-r^2) \cr
  1-r^2 & -\left( e^{+2i\phi}-r^2 e^{-2i\phi} \right)
 \end{pmatrix},
\end{align}
where $\phi$ and $r$ are defined by $1/k_A \xi_B \equiv \tan \phi$ 
($E=V\cos^2\phi$) and $r\equiv e^{-b/\xi_B}$.
Thus, if $2\phi$ and $r$ are identified with $\omega l/v_c$ and the EMP coupling strength, respectively,
there is a close correspondence between the two systems: for example, 
studying the EMP which has approximately $\omega_s$ ($2\omega_s$) is the same thing as 
studying the electron with $\phi\to \pi/2$ ($\pi$) near the bottom (top) of the potential energy $E\to 0$ ($E\to V$).
More explicitly, 
the above derivation of the $T(\omega)$ matrix originates from the fact that $T_{AB}$ is rewritten as
\begin{align}
 T_{AB} = \frac{\sqrt{k_A^2+\xi_B^{-2}}}{2k_A} 
 \begin{pmatrix}
  e^{-i\phi} & e^{i\phi} \cr
  e^{i\phi} & e^{-i\phi}
 \end{pmatrix}
 = 
 \frac{\sqrt{k_A^2+\xi_B^{-2}}}{2k_A} 
 \begin{pmatrix}
  1 & 1 \cr
  e^{i\phi} & e^{-i\phi}
 \end{pmatrix}
 \begin{pmatrix}
  e^{-i\phi} & 0 \cr
  0 & e^{i\phi}
 \end{pmatrix},
\end{align}
in terms of $\phi$ defined by $k_A + i/\xi_B = \sqrt{k_A^2 + \xi_B^{-2}} e^{i\phi}$
[see Eqs.~(\ref{eq:TABTBA}) and (\ref{eq:Tmat})].
This similarity is more than what is naturally expected from the point of view of wave mechanics
and suggests the physical phenomena observed in a superlattice may manifest itself in EMP crystals.

By comparing Eq.~(\ref{eq:KPR}) with Eq.~(\ref{eq:A}),
we find that the superlattice has direct relevance to the EMP chain if we assume that 
\begin{align}
 & \cosh(b/\xi_B) = \frac{1+r^2}{2r}, \\
 & \frac{1}{k_A \xi_B} - k_A \xi_B = 2\cot\left( \frac{\omega \ell}{v_c} \right).
\end{align}
The former equation confirms $r= e^{-b/\xi_B}$,
and a small decay length ($\xi_B \ll b$) caused by a large $V$ 
corresponds to a weak coupling.
Meanwhile, a large decay length ($\xi_B \gg b$) corresponds to a strong coupling, 
which seems to be a reasonable correspondence.
The latter equation leads to 
$1/k_A \xi_B = -\tan \left( \frac{\omega \ell}{2v_c} \right)$.
The minus sign just appears as a result of 
the correspondence between Eq.~(\ref{eq:UTRTRU}) and Eq.~(\ref{eq:URTU}) for positive $\omega$.

The correspondence $r= e^{-b/\xi_B}$, where $\xi_B$ is energy dependent while $r$ is just a coupling constant, 
is not easy to understand. 
Such an apparent disagreement may be hidden by taking the limit of $V\to \infty$ and $b\to 0$ 
in such a way that $Vb$ is a constant, namely the Dirac delta potential.
Though $b/\xi_B^2$ is a nonzero constant, $b/\xi_B \to 0$, and the potential corresponds to 
the strong coupling limit ($r\to 1$) of an EMP chain.
On the other hand, it has been shown that $e^{-b/\xi_B}$ can be used as a small parameter in perturbation theory to 
solve the Kronig-Penney model (and to extract a tight-binding parameter, such as hopping integrals).~\cite{Marsiglio2017}  
Thus, EMP chains may cover the Kronig-Penney model with various unit-cell structures.
We note that the bound states caused by a negative Dirac delta potential
can be studied by taking the $a\to 0$ limit and changing the origin of the energy as $E\to E+V$,
for the localized states in layer B ($E<0$).
For the bound states, $k_A^2 a$ is a constant and 
$k_B=i/\xi_B$, where $\xi_B^{-1}=\sqrt{2m|E|}/\hbar$.
In this case, $r\equiv e^{-b/\xi_B}$ may take a general value.
In a bipartite model, such a bound state can be doubled in the unit cell, 
which has been examined from the point of view of toplogically protected edge states.~\cite{Smith2019}
Such a model is more relevant to an EMP chain containing two domains with different domain sizes $L_1$ and $L_2$
(or different coupling strength $r_1$ and $r_2$) in the unit cell.

 \bibliography{/Users/Sasaki/tex/bib/library}

\end{document}